\documentclass[10pt]{article}
\usepackage{amsthm,amsfonts,amsmath}

\title{Fluctuations of the maximal particle energy\\
of the quantum ideal gas\\ and random partitions}
\author{A.~Vershik, Yu.~Yakubovich}

\newcommand{\CP}{\mathcal{P}}
\newcommand{\BC}{\mathbb{C}}
\newcommand{\BD}{\mathbb{D}}
\newcommand{\BR}{\mathbb{R}}
\newcommand{\bb}{{\boldsymbol{b}}}
\newcommand{\la}{\lambda}

\theoremstyle{plain}
\newtheorem{theorem}{Theorem}

\theoremstyle{remark}
\newtheorem*{remark*}{Remark}
\newtheorem*{example*}{Example}

\newtheorem*{definition*}{Definition}

\begin{document}
\maketitle

\begin{abstract}
We investigate the limiting distribution of the fluctuations of the maximal
summand in a random partition of a large integer with respect to a
multiplicative statistics. We show that for a big family of Gibbs
measures on partitions (so called generalized Bose--Einstein statistics)
this distribution is the well-known Gumbel distribution which usually
appears in the context of indepedent random variables.
In particular, it means that the (properly rescaled) maximal energy
of an individual particle in the grand canonical
ensemble of the $d$-dimensional quantum ideal gas has the Gumbel distribution
in the limit.

We also apply our result to find the fluctuations of the height of a random
3D Young diagram (plane partition) and investigate the order statistics
of random partitions under generalized Bose--Einstein statistics.
\end{abstract}

\section{Introduction}
The link between the energy distribution in grand canonical ensembles of a quantum ideal gas and
the asymptotic behavior of random partitions of integers was established in several papers. For
example, the problem about the limit shape of the energy distribution in such ensemble was
considered in \cite{Vershik-FA96,Vershik-UMN97}. In this paper we are interested in a more subtle
question---the limit distribution of the maximal energy of a particle in the grand canonical
ensemble of the ideal Bose gas of dimension $d$ with the Gibbs statistics. We give the answer (see
Theorem~\ref{thm-max-gas} below), but the question is involved in a general context which has some
history.

Consider the set of all partitions $\CP =\cup \CP(n)$ of all natural numbers
$n$ with the so called multiplicative (Gibbs) statistics,---it means that,
firstly,  occupation numbers are independent with respect to this
statistics and, secondly, restrictions of the statistics on the subset of
all partitions of the given number $n$ does not depend on the parameter of
the Gibbs measure (activity). We give precise definition below.
The simplest example of a multiplicative
statistics is the so called Poissonization of the uniform measures on
partitions of integers;
a more general example is the Gibbs measures of the quantum ideal gas, either
Fermi or Bose (see \cite{Vershik-FA96}). It is possible (under some special
conditions) to find the limit shape of partition, i.e.\ the distribution of
energy \cite{Vershik-UMN97}.
The next problem is to consider the distribution of the fluctuations. In the
bulk of summands the fluctuations have a Gaussian distribution (see
\cite{Vershik-Freiman-Yakubovich,Yak-99}). But the fluctuations on the
``ends'' of partition, i.e.\ the
fluctuations of the maximal summands, have completely different form.
We can compare this situation to the same question for Plancherel measure
where the distribution of the fluctuations of the longest increasing
subsequences (or the maximal row of a random Young diagram) with respect to
the Plancherel measure is the Tracy--Widom distribution
(\cite{Baik-Deift-Johansson}).
For our problem we get a so called \textit{Gumbel distribution}
with the distribution function $e^{-e^{-t}}$ which occurs also
in the theory of extreme values of independent random variables (see,
e.g.,~\cite{embrechtsetal}), but its
appearance in our case has a different nature.

First time this distribution appeared in similar problems in the pioneering
work of Erd\H{o}s and Lehner \cite{Erdos-Lehner} where they proved
that the maximal summand in a random uniformly distributed partition of a large
positive integer $n$ is approximately
$(\sqrt{6n}\log n)/(2\pi)$ and found a right scaling
and a limiting distribution for the maximal summand $m(\la)$ in a random
partition $\la$ if the partition is taken uniformly among all partitions of
$n$. Exactly, their result is that for all $x\in\BR$
\[
\lim_{n\to\infty}\frac{1}{p(n)}\#\Bigl\{
\la:\frac{m(\la)-(\sqrt{n}\log n)/(2c)}{\sqrt{n}}\le x\Bigr\}
=e^{-c^{-1}e^{-cx}},\quad c=\frac{\pi}{\sqrt{6}},
\]
where $p(n)$ is
the number of all partitions of $n$.
The proof presented in \cite{Erdos-Lehner} is based on some
combinatorial estimates for numbers of partitions with certain properties,
asymptotic relations based on these estimates and the
inclusion-exclusion principle.

Similar behavior is known for other measures on partitions, for example,
for the uniform measure on set partitions. Namely, take at random a
partition of the set $\{1,\dots,n\}$ and look only on block sizes, which give
a (nonuniform) distribution on partitions of an integer $n$. In this case
the limiting distribution of the maximal block size is a discrete approximation
to the Gumbel distribution, see~\cite{Sachkov}, Section~IV.5.

In this note we present a different approach to this problem allowing us
to generalize this result to a more general family of measures on partitions.
We use Poissonized measures and prove all our results
only for them.  These results for Poissonized measures can be
extended to the corresponding statements for original measures.  General
principle allowing to do it will be presented in another paper.

Now we give the precise description of what is called multiplicative
statistics.
Let $\CP(n)$ be the set of partitions of an integer $n$
and $\CP=\cup_{n\ge0}\CP(n)$
be the set of all partitions, as above.
For $\la=(\la_1,\dots,\la_\ell)\in\CP$,
$\la_1\ge\dots\ge\la_\ell>0$,
we write $r_k(\la)=\{i:\la_i=k\}$ for the
number of summands $k$ in partition $\la$,
so called \textit{$k$-th occupation number}.
Consider a sequence of functions $f_k(z)$, $k\ge1$, analytic in the open disk
$\BD=\{z\in\BC:|z|<R\}$, $R=1$ or $R=\infty$, such that $f_k(0)=1$
and assume that (\textit{i}) the Taylor series
\begin{equation}\label{fkofz}
f_k(z)=\sum_{j=0}^\infty s_{k}(j)z^j
\end{equation}
have all coefficients $s_k(j)\ge0$ and (\textit{ii}) the infinite product
\begin{equation}\label{deffofx}
F(x)=\prod_{k=1}^\infty f_k(x^k)
\end{equation}
converges in $\BD$.  Then one can define a family of probability measures
$\mu_x$, $x\in(0,R)$, on
the set of all integer partitions $\CP$
in the following way: put
\begin{equation}\label{muxofrkeqj}
\mu_x\bigl(\{\la\in\CP:r_k(\la)=j\}\bigr)=\frac{s_k(j)x^{kj}}{f_k(x^k)},
\end{equation}
and assume that different occupation numbers are independent.  Note that
in order to specify measure $\mu_x$  it suffices to fix $F(x)$ along
with its decomposition~\eqref{deffofx}. At the same time just specifying
$F(x)$ is not enough.

\begin{definition*} The family of measures $\mu_x$ defined
by~\eqref{muxofrkeqj}
and satisfying conditions~(\textit{i}) and~(\textit{ii}) is called a
\textit{multiplicative} family of measures (or a \textit{multiplicative
statistics}) on partitions.  In this case we say that the statistics is
determined by decomposition~\eqref{deffofx}.
\end{definition*}

This notion in a general context was introduced in~\cite{Vershik-FA96};
similar technique was exploited earlier by Fristedt~\cite{Fristedt}
for the uniform measure on partitions.
The key feature of these measures is the fact that conditional probability
measures induced on
$\CP(n)$ do not depend on $x$, for all $n$.  We denote these conditional
measures by $\mu^n=\mu_x\big|_{\CP(n)}$.
Thus measures $\mu_x$ can be considered
as a Poissonization of measures $\mu^n$ which makes random variables $r_k$
independent.  Many statements which hold for $\mu_x$ hold for $\mu^n$ too,
see, for instance, \cite{Vershik-FA96,Vershik-Dembo-Zeitouni};
at the same time, the investigation of measures $\mu_x$ is simpler because
$r_k$ are independent.

We call the family $(\CP(n),\mu^n)$ \textit{small canonical ensemble of
partitions}\footnote{The notion \textit{microcanonical ensemble} is usually
used for the set of partitions $n=\la_1+\dots+\la_\ell$
with both weight~$n$ and length~$\ell$ fixed,
so we introduce the term small canonical ensemble for an ensemble with fixed
energy but varying number of particles.} and the family
$(\CP,\mu_x)$---\textit{grand canonical ensemble of
partitions}, in view of similarities with statistical physics,
see~\cite{Vershik-FA96}.

We restrict our attention to the special case of a so-called
generalized Bose--Einstein statistics defined by~\eqref{muxofrkeqj} with
$f_k(z)=1/(1-z)^{b_k}$.
Thus equation~\eqref{deffofx} specializes to
\begin{equation}\label{Fboseeinstein}
F_{\bb}(x)=\prod_{k=1}^\infty\frac{1}{(1-x^k)^{b_k}},
\end{equation}
$\bb=\{b_k\}$.
More exactly, first we treat only the case when
$b_k=ck^\beta$ for some $c>0$ and $\beta>-1$.  In this case direct
calculations can be made to show that the limiting distribution of a maximal
summand is the Gumbel distribution, as done in Theorem~\ref{maxbkthm}.  Note
that the Poissonization of the uniform measure on partitions belongs to this
family (all $b_k=1$ in this case).  For the uniform measure the distribution
of a maximal summand coincides with the distribution of the length of
a partition (that is the number of summands), but in a general case these
distributions are essentially different.

For integer $b_k$ these partitions are often called
\textit{colored partitions} because the measure on the small canonical
ensemble is induced by the uniform measure on partitions with additional
structure: each summand~$k$ can be colored in one of~$b_k$ colors, and two
partitions are identical when number of summands of each size and color
coincide, see~\cite{Andrews}.  However while the requirement that $b_k$ are
integers is natural from the combinatorial point of view, it is not needed
analytically and we consider all real positive $b_k$.

Our main result is presented in Section~\ref{sec-gas}.  It concerns the
distribution of the maximal energy of an individual particle in the large
canonical ensemble of the $d$-dimensional quantum ideal gas (see, e.g.,
\cite{Huang,Landau-Lifshitz} for background).  We show that
\textit{after a suitable scaling the limiting distribution of the maximal
energy of an individual particle is the Gumbel distribution}.  The exact
statement is given in Theorem~\ref{thm-max-gas}.  Note that while the grand
canonical ensemble of the quantum ideal gas is the partial case of the
generalized Bose--Einstein statistics, Theorem~\ref{maxbkthm} can not be
directly applied since $b_k$ do not have a form $ck^\beta$ even
asymptotically.

In Section~\ref{sec-3d} we sketch the application of our result to another
example, namely to the the distribution of the height of 3D Young diagrams.
As above, we consider only the Gibbs
measure on these objects.  In Section~\ref{sec-small-ens} we conjecture the
limiting behavior in small canonical ensembles of partitions.  The last
section is devoted to the investigation of the order statistics, or, in
other words, of the spacing between first largest summands in partition.

\section{Maximal summand}\label{sec-max}
We consider the special case of multiplicative statistics, namely measures
$\mu_{\bb,x}$ defined by~\eqref{muxofrkeqj} with
$f_k(z)=1/(1-z)^{b_k}$, $\bb=\{b_k\}=\{ck^\beta\}_{k\ge1}$ with
$\beta>-1$. The convergence radius of $f_k$ and $F$ is $R=1$.
We denote the maximal summand in a
partition~$\la=(\la_1,\dots,\la_\ell)$, $\la_1\ge\dots\ge\la_\ell>0$,
by~$m(\la)\equiv\la_1$.

\begin{theorem}\label{maxbkthm}
Let the measure $\mu_{\bb,x}$ be defined by~\eqref{muxofrkeqj} where
\[
F(x)=\prod_{k\ge1}{f_k(x^k)}=\prod_{k\ge1}\frac{1}{(1-x^k)^{ck^\beta}}
\]
for some $c>0$ and $\beta>-1$. Then, for all $t\in\BR$
\begin{equation}\label{max-be}
\lim_{x\to1}\mu_{\bb,x}\bigl\{\la\in\CP:m(\la)(1-x)-A(x)\le t\bigr\}
=e^{-e^{-t}},
\end{equation}
where
\begin{equation}\label{be-aofx}
A(x)=(\beta+1)\bigl|\log(1-x)\bigr|+\beta\log\bigl|\log(1-x)\bigr|
+\beta\log(\beta+1)+\log c.
\end{equation}
\end{theorem}

\begin{remark*} The standard form of a limit theorem is to find
$\lim\mu_x\bigl\{\la\in\CP:\tfrac{m(\la)-a(x)}{b(x)}\le t\bigr\}$.
Expression~\eqref{max-be} looks differently but can be rewritten in this
form taking $a(x)=A(x)/(1-x)$ and $b(x)=1/(1-x)$.
\end{remark*}

\begin{proof} We shall not just verify the statement of the Theorem but
also show how it can be deduced.  First, note that the
probability $\mu_{\bb,x}\{\la\in\CP:m(\la)\le M\}$ tends to 0 for fixed $M$
as $x\to R$, so in order to
get sensible results we should take $M$ depending on $x$.  More exactly,
in order to get the limit theorem in form~\eqref{max-be} we take
\begin{equation}\label{defmxt}
M=M(x,t)=a(x)+b(x)t,
\end{equation}
where $a$ and $b$ are functions of $x$ presumably
growing to infinity as $x\to1$, and $t$ is a parameter.  Since we want
$M(x,t)$ grow to infinity for all fixed $t$, it follows that $b(x)=o(a(x))$
as $x\to1$.

Since measures $\mu_{\bb,x}$ are more
adjusted to work with occupation numbers $r_k$, we reformulate the
question in terms of $r_k$ in the following way:
for any $M\ge1$
\begin{equation}\label{mumaxlem}
\mu_{\bb,x}\{\la\in\CP:m(\la)\le M\}
=\mu_x\{\la:r_k(\la)=0\text{ for }k>M\}
=\prod_{k>M}\frac{1}{f_k(x^k)}.
\end{equation}
Thus we just have to find functions $a(\cdot)$ and $b(\cdot)$ such that,
as $x\to R$,
the product of $1/f_k(x^k)$ taken for $k>M(x,t)=a(x)+b(x)t$ tends to some
non-degenerate distribution function.

We take minus logarithm of \eqref{mumaxlem} to conclude that
\[
-\log(\mu_{\bb,x}\{\la:m(\la)\le M(x,t)\})=-\sum_{k>M(x,t)}b_k\log
(1-x^k).
\]
Let us assume that
\begin{equation}\label{assumM}
M(x,t)>p|\log(1-x)|/|\log x|
\end{equation}
for some $p>0$; under this assumption we have $0\le\sup_{k\ge m(x,t)}x^k\le
C(1-x)^p$ for some $C>0$.  So, using \eqref{assumM} and putting
$b_k=ck^\beta$ we can calculate the sum above explicitly:
\begin{multline}\label{estlogmux}
-\log(\mu_{\bb,x}\{\la:m(\la)\le M(x,t)\})
=\sum_{k>M(x,t)}ck^\beta x^k+O\Bigl(\sum_{k>M(x,t)}k^\beta x^{2k}\Bigr)\\
=\frac{c\,(M(x,t))^\beta\, x^{M(x,t)}}{1-x}\bigl(1+O(1/M(x,t))\bigr)
\end{multline}
as $x\to1$ with fixed $t$.
We take a logarithm once more and arrive to
\begin{multline}\label{loglogmux}
\log\bigl(-\log(\mu_{\bb,x}\{\la:m(\la)\le M(x,t)\})\bigr)\\
=-M(x,t)|\log x|+|\log(1-x)|+\beta\log M(x,t)+\log c+O\bigl(1/M(x,t)\bigr)
\end{multline}
as $x\to1$.

We are seeking $M(x,t)$ in the form~\eqref{defmxt}, and taking
$b(x)=1/|\log x|$ in~\eqref{defmxt} we can
rewrite~\eqref{loglogmux} as
\begin{multline*}
\log\bigl(-\log(\mu_{\bb,x}\{\la:m(\la)\le M(x,t)\})\bigr)\\
=-a(x)|\log x|+|\log(1-x)|+\beta\log a(x)+\log c-t+O\bigl(1/a(x)\bigr).
\end{multline*}
So if we could find $a(x)$ such that
\begin{equation}\label{maxbktag1}
-a(x)|\log x|+|\log(1-x)|+\beta\log a(x)+\log c=0
\end{equation}
then $\mu_{\bb,x}\{\la:m(\la)\le M(x,t)\}$ would tend to $e^{-e^{-t}}$ for
fixed $t$ as $x\to1$.
We are searching the solution of~\eqref{maxbktag1} in a form
\[
a(x)=\frac{(\beta+1)|\log(1-x)|+a_1(x)}{|\log x|}
\]
where
$a_1(x)=o(|\log(1-x)|)$. After a substitution into~\eqref{maxbktag1} we
immediately see that $a_1(x)=\beta\log|\log(1-x)|+\log c+o(1)$.  It remains
to use the relation $|\log x|\sim 1-x$ as $x\to1$ to get~\eqref{max-be}.

Now note that for this choice of $a(\cdot)$ and $b(\cdot)$, the
assumption~\eqref{assumM} is satisfied iff $\beta>-1$, which justifies all
computations above.
\end{proof}

\section{The quantum ideal gas}\label{sec-gas}
Here we apply our results to obtain the limiting distribution of the maximal
energy of an individual particle in the quantum ideal gas in $\BR^d$.
(The most interesting case is of course
$d=3$ but the method works for all $d\ge1$).  We describe briefly the
connection of the quantum ideal gas to the partition theory; for a
detailed exposition see, e.g.,~\cite{Huang,Landau-Lifshitz}.
To each configuration in the phase space there corresponds a partition
of an integer $n$: in the suitable units, the energies of individual
particles become summands in the partition
and the energy of the whole system becomes $n$. There are
$j_d(k)=\#\bigl\{(k_1,\dots,k_d)\in\mathbb{Z}^d:k_1^2+\dots+k_d^2=k\bigr\}$
distinct positions of particles in the phase space such that
a particle in these positions has (rescaled) energy $k$.
Consequently, the measure on partitions induced by the grand canonical
Gibbs measure on the quantum $d$-dimensional ideal bosonic gas is a
multiplicative
measure determined by the decomposition~\eqref{Fboseeinstein} with
$b_k=j_d(k)$. This observation was used
by Vershik~\cite{Vershik-FA96,Vershik-UMN97} to find a distribution
of energy among particles in these settings.  The following statement
describes the behavior of the maximal energy of an individual particle.

\begin{theorem}\label{thm-max-gas} Let measure $\mu_{\bb,x}$ be the
multiplicative measure on partitions originating from the quantum ideal
Bose gas, i.e.\ it is defined by~\eqref{muxofrkeqj} where
\[
F(x)=\prod_{k\ge1}f_k(x^k)=\prod_{k\ge1}\frac{1}{(1-x^k)^{j_d(k)}}.
\]
Then, for all $t\in\BR$,
\[
\lim_{x\to1}\mu_x\{\la\in\CP:m(\la)(1-x)-A(x)\le t\}=e^{-e^{-t}}
\]
\textit{where}
\[
A(x)=\tfrac{d}{2}\bigl|\log(1-x)\bigr|+\tfrac{d-2}{2}\log\bigl|\log(1-x)\bigr|
+\tfrac{d}{2}\log\tfrac{d}{2}+\log\tfrac{\pi^{d/2}}{\Gamma(d/2+1)}.
\]
\end{theorem}

Classical results on $j_d(k)$ state that
$j_d(k)=O(k^{d/2-1})$ for $d\ne4$ and $j_4(k)=O(k\log k)$, see
\cite{Grosswald}. Nevertheless,  $j_d(k)k^{1-d/2}$ has no limit.
Thus, Theorem~\ref{maxbkthm} can not be directly applied to obtain results
on limiting behavior of the maximal energy of an individual particle.
But the ideas used in its proof still work.

\begin{proof}[Proof of Theorem~\ref{thm-max-gas}]
We are going to get estimate analogous to~\eqref{estlogmux} used in
the proof of Theorem~\ref{maxbkthm}, which was crucial to get that result.
Once we have such estimate the rest of the proof almost literally repeats
the proof of Theorem~\ref{maxbkthm}.

Take $M(x,t)=\bigl[(A(x)+t)/|\log x|\bigr]$; then  $\sup_{k>M(x,t)}x^k\to0$.
As in the proof of Theorem~\ref{maxbkthm} we obtain
\begin{equation}\label{gasthmt1}
-\log(\mu_{\bb,x}\{\la:m(\la)\le M(x,t)\})
=\sum_{k>M(x,t)}j_d(k)x^k+O\Bigl(\sum_{k>M(x,t)}j_d(k)x^{2k}\Bigr).
\end{equation}
Both sums above can be treated in a similar way.  We show that the first sum
tends to $e^{-e^{-t}}$ as $x\to1$; replacing $x$ by $x^2$ in suitable places
shows that $O(\cdot)$ vanishes.

Denote $J_d(k)=\sum_{i\le k} j_d(i)$.  It is well known
(see~\cite{Grosswald,Valfish}) that $J_d(k)=C_dk^{d/2}+E_d(k)$ where
$C_dk^{d/2}=\tfrac{\pi^{d/2}}{\Gamma(d/2+1)}k^{d/2}$ is the volume of a
$d$-dimensional ball of radius~$\sqrt{k}$ and the error term
$E_d(k)=O(k^{\alpha_d})$ where $\alpha_d$ can be taken as follows:
$\alpha_1=0$, $\alpha_2=1/3$, $\alpha_3=3/4$, $\alpha_4=1+\delta$
($\delta>0$) and $\alpha_d=d/2-1$ for $d\ge5$ (better estimates are known
but we do not need them).

For any $M\ge1$ we can write
\begin{multline*}
\sum_{k\ge M} j_d(k)x^k=\sum_{k\ge M} (J_d(k)-J_d(k-1))x^k\\
=C_d\sum_{k\ge M}\bigl(k^{d/2}-(k-1)^{d/2}\bigr)x^k
+(1-x)\sum_{k\ge M} E_d(k)x^k - E_d(M-1)x^M\\
=C_d\frac{d}{2}\,\frac{x^MM^{(d-2)/2}}{|\log x|}+R_d(M),
\end{multline*}
where the error term~$R_d(M)$ can be estimated as follows:
$|R_d(M)|<|R'_d(M)|+|R''_d(M)|$ with
\begin{align*}
R'_d(M)&{}=C_d\sum_{k\ge M}\bigl(k^{d/2}-(k-1)^{d/2}\bigr)x^k-
C_d\frac{d}{2}\,\frac{x^MM^{(d-2)/2}}{|\log x|},\\
R''_d(M)&{}=(1-x)\sum_{k\ge M} E_d(k)x^k - E_d(M-1)x^M.
\end{align*}
Using the integral approximation for the sum one can find that
$R'_d(M)\le K_1x^MM^{(d-2)/2}$
and estimates on $E_d$ show that $R''_d(M)\le K_2x^MM^{\alpha_d}$ for some
$K_1,K_2>0$.
Taking $M=M(x,t)=\bigl[(A(x)+t)/|\log x|\bigr]$ we see that the leading
error term is $R''_d(M)$ and that for fixed~$t$ the
inequality $|R_d(M(x,t))|<K(1-x)^{d/2-\alpha_d}$ holds for some $K>0$. It
remains to check that our choice of $M(x,t)$ implies that
\[
\lim_{x\to1}C_d\frac{d}{2}\,
\frac{x^{M(x,t)}\bigl(M(x,t)\bigr)^{(d-2)/2}}{|\log x|}\to e^{-e^{-t}}.
\]
Almost the same calculation verifies that the argument of $O(\cdot)$
in~\eqref{gasthmt1} vanishes as~$x\to1$.
\end{proof}

One can also consider the Fermi--Dirac $d$-dimensional quantum ideal gas,
which also induces the multiplicative statistics on partitions with
$f_k(z)=(1+z)^{j_d(k)}$.  In the grand canonical ensemble it
has exactly the same limiting behavior of the
maximal energy of a particle.  It can be seen analytically since the
limiting behavior of the maximal energy depends only on the first Taylor
coefficients of functions~$f_k$, and they are the same for the Fermi--Dirac
and Bose--Einstein statistics.  It worse noticing that most of statistical
properties of Fermi and Bose gases are different.

\section{A section of 3D Young diagram}\label{sec-3d}
Let us consider a set of 3D Young diagrams (or plane partitions) of weight~$N$.
A \textit{3D Young diagram}
is a $\mathbb{Z}_+$-valued function $h(u,v)$ of two arguments
$u,v\in\mathbb{R}_+$ such that it has a finite support,
is non-increasing in both arguments
and if it is discontinuous in a point $(u,v)$ then either $u$ or $v$ is
integer.  A \textit{weight} of a diagram $h(u,v)$ is
\[
N(h(\cdot,\cdot))=\iint_{\BR_+^2}h(u,v)\,du\,dv
\]
and it is obviously an integer.  The graph of $h(u,v)$ in $\BR^3$
is an upper bound of a set which can be constructed from unit cubes in
the same way as ordinary 2D Young diagram is constructed from unit
squares (or boxes); this analogy explains the name of these objects.
We denote the set of all 3D Young diagrams by $\CP_{3D}$.

Given a number $x\in(0,1)$ one can consider a probability measure on the
set of all 3D Young diagrams with probability of any diagram $h(\cdot,\cdot)$
is proportional to $x^{N(h(\cdot,\cdot))}$. The generating function
for numbers $p_3(N)$ of 3D Young diagrams of weight~$N$ is well known:
\[
\sum_{N\ge0}p_3(N)x^N=\prod_{k\ge1}\frac1{(1-x^k)^k},
\]
see, e.g.,~\cite{Andrews}. The existence of a limit shape for
3D Young diagrams was proved by the first author; the exact formulas were
found later by
Kenyon and Cerf~\cite{Kenyon-Cerf} using a variational technique and
further investigated by
Okounkov and Reshetikhin~\cite{Okounkov-Reshetikhin}.
Thus any section of a graph of 3D Young diagram has a limit shape also.
In particular, a section by the plane $u=v$ can be considered as the Young
diagram of an ordinary partition (up to a factor $\sqrt{2}$).
It turns out that a distribution of these diagram
will be exactly the distribution defined by~\eqref{Fboseeinstein}
with $b_k=k$ (we should treat $h$ axis in the 3D case as $t$ axis in 2D).
The correspondence between colored partitions and diagonal sections of 3D
Young diagrams is rather complicated and includes Bender--Knuth
bijection between random infinite integer matrices and pairs of
semi-standard Young tableaux and the correspondence between these pairs and
3D diagrams.
It was introduced by the first author in his talk~\cite{Vershik-Vienna},
its detailed description and consequences will be presented in a separate
paper.

Thus, direct application of Theorem~\ref{maxbkthm} leads to the following
result:
\begin{theorem}\label{max3dthm} Let $\mu_x$ be the Gibbs probability measure
on $\CP_{3D}$, i.e.\ the measure of a particular diagram
$h(\cdot,\cdot)$ is proportional to $x^{N(h(\cdot,\cdot))}$. Then,
for all $t\in\BR$,
\begin{equation}\label{max3d}
\lim_{x\to1}\mu_x\bigl\{h(\cdot,\cdot)\in\CP_{3D}:
  h(0,0)(1-x)-A(x)\le t\bigr\}
=e^{-e^{-t}},
\end{equation}
where
\begin{equation}
A(x)=2\bigl|\log(1-x)\bigr|+\log\bigl|\log(1-x)\bigr|
+\log 2.
\end{equation}
\end{theorem}

\section{Statements for small canonical ensemble}\label{sec-small-ens}
The corresponding results in small canonical ensemble of partitions,
that is results for measures $\mu^n$,
can be obtained by taking $x=x(n)$ depending on $n$ so that the expected
weight of partition ($\sum_k kr_k$) is $n$.
For the measures considered in
Theorem~\ref{maxbkthm} it can be achieved by taking
\[
x=x(n)=1-\sqrt[\beta+2]{\frac{c\Gamma(\beta+2)\zeta(\beta+2)}{n}}
\]
($\zeta$ is the Riemann zeta function), see~\cite{Vershik-FA96}.
We say that grand canonical and small canonical ensembles
\textit{are equivalent} for some functional $G$ on partitions
if the distributions of the functional~$G$ w.r.t.\ $\mu^n$ and w.r.t.\
$\mu_{x(n)}$ are asymptotically the same.  It seems that
the ensembles are equivalent for the functional
of rescaled maximal summand in partition. We
shall return to this question in another paper.

In the assumption that ensembles are equivalent Theorem~\ref{maxbkthm}
yields the following result:
\textit{if measures $\mu^n$ are conditional probability measures
induced on $\CP(n)$ by the generalized Bose--Einstein measures $\mu_x$
determined by decomposition~\eqref{Fboseeinstein} with $b_k=ck^\beta$, then}
\begin{equation}\label{limitmun}
\lim_{n\to\infty}\mu^n\Bigl\{\la\in\CP(n):
\sqrt[\beta+2]{\frac{c\Gamma(\beta+2)\zeta(\beta+2)}{n}}\,m(\la)
-A_n\le t\Bigr\}
=e^{-e^{-t}},
\end{equation}
\textit{where}
\begin{multline*}
A_n=\frac{\beta+1}{\beta+2}\log n+\beta\log\log n\\
+\beta\log\frac{\beta+1}{\beta+2}
-\frac{\beta+1}{\beta+2}\log\bigl(\Gamma(\beta+2)\zeta(\beta+2)\bigr)
+\frac{1}{\beta+2}\log c.
\end{multline*}
The terms in the second line above do not depend on $n$ so they constitute a
constant correction term, while the terms in the first line show how
maximal summand is growing with the growth of~$n$.

In two particular examples considered above, the exact computation can be
made.  For the case of the quantum ideal gas of total energy~$n$ (in the
suitable units so that $n$ is integer), we should take
\[
x=x(n)=1-\sqrt[d/2+1]{\frac{d\pi^{d/2}\zeta(d/2+1)}{2n}}
\]
in the grand canonical ensemble to get the best approximation of
the small canonical measure~$\mu^n$. Thus, under the assumption of
equivalence of ensembles, \textit{in the small canonical ensemble of
$d$-dimensional quantum ideal Bose gas}
\[
\lim_{n\to\infty}\mu^n\Bigl\{\la\in\CP(n):
m(\la)\Bigl(\tfrac{d\pi^{d/2}\zeta(d/2+1)}{2n}\Bigr)^{2/(d+2)}
  -A(n)\le t\Bigr\}=e^{-e^{-t}}
\]
\textit{with}
\begin{multline*}
A_n=\tfrac{d}{d+2}\log n+\tfrac{d-2}{d}\log\log n\\
+\tfrac{d^2}{2(d+2)}\log\tfrac{d}{2}+\tfrac{d-2}{2}\log\tfrac{2}{d+2}
-\tfrac{d}{d+2}\log(\zeta(d/2+1)/\pi)-\log\Gamma(d/2+1).
\end{multline*}

Similarly, it follows from~\eqref{limitmun}
that a height of a typical 3D Young diagram (i.e.\ $h(0,0)$) of weight~$N$
behaves as $\frac{2^{2/3}}{3\zeta(3)^{1/3}}N^{1/3}\log N$
as $N\to\infty$. More exactly, under the assumption of equivalence of
ensembles,
\textit{denoting by $\nu^N$ the uniform
measure on 3D Young diagrams of weight $N$, we have}
\[
\lim_{N\to\infty}\nu^N\Bigl\{h:\tfrac{(2\zeta(3))^{1/3}}{N^{1/3}}h(0,0)
-\tfrac23\log N-\log\log N
-\log\tfrac23+\tfrac23\log 2\zeta(3)\le t\Bigr\}=e^{-e^{-t}}.
\]

\section{Order statistics}
Now let us consider the upper order statistics of random partitions, that is
the sequence of the first $d$ largest summands $m_1(\la)\ge\dots\ge m_d(\la)$
in a random partition $\la$.  We consider only the settings of
Theorem~\ref{maxbkthm}; generalizations to other measures considered above
are straightforward.

\begin{theorem}\label{ordthm}
Let measures~$\mu_{\bb,x}$ be defined by~\eqref{muxofrkeqj} with
$f_k(z)=1/(1-z)^{b_k}$, $\bb=\{b_k\}=\{ck^\beta\}_{k\ge1}$ and
$\beta>-1$.  Let $A(x)$ be defined by~\eqref{be-aofx}.
Then the distribution of rescaled upper order statistics
$\bigl(m_i(\la)(1-x)-A(x)\bigr)_{i=1,\dots,d}$ converges weakly to the
distribution on $\BR^d$ with the joint density
$\exp\bigl(-e^{-t_d}-\sum_{i=1}^d t_i\bigr)$ for $t_1>\dots>t_d$ and zero
otherwise.
\end{theorem}

\begin{proof}
Let us fix $t_1>t_2>\dots>t_d$ and let $\delta>0$ be such that
intervals $\Delta_i=[t_i,t_i+\delta]$ are disjoint.  Then the probability
that each $m_i(\la)$ after rescaling gets in the interval $\Delta_i$ is
\begin{equation}\label{mstag2}
\mu_x\{\la\in\CP:m_i(\la)(1-x)-A(x)\in\Delta_i,i=1,\dots,d\}
=\frac{\prod_{i=1}^d S_i}{\prod_{k>\frac{A(x)+t_d}{1-x}}f_k(x^k)}
\end{equation}
where
\[
S_i=\sum_{k\in[\frac{A(x)+t_i}{1-x},\frac{A(x)+t_i+\delta}{1-x}]} s_k(1)x^k.
\]
(Recall that $s_k(1)=b_k=ck^\beta$ and $s_k(1)x^k/f_k(x^k)$ is the
probability that $r_k=1$.) Note that
we estimated the denominator in the RHS of
\eqref{mstag2} while we proved Theorem~\ref{maxbkthm} (see
equations~\eqref{loglogmux} and~\eqref{maxbktag1}) and that it
tends to $1/\exp(-e^{-t_d})$ as $x\to1$.  Let us
estimate the numerator, that is sums~$S_i$. In view of~\eqref{estlogmux},
\[
S_i=\frac{c}{1-x}\bigl(\bigl(M(x,t_i)\bigr)^\beta x^{M(x,t_i)}
 -\bigl(M(x,t_i+\delta)\bigr)^\beta x^{M(x,t_i+\delta)}\bigr)
\]
where $M(x,t)=\frac{A(x)}{1-x}+\frac{1}{1-x}t$.
This choice of $M(x,t)$ implies that $S_i\to e^{-t_i}(1-e^{-\delta})$ as
$x\to1$ for fixed~$t_i$.  Taking $\delta\to0$ limit finishes the proof.
\end{proof}

\begin{remark*} The same behavior of upper order statistics is known for
samples of $n$ i.i.d.\ random variables lying in the attraction domain of
Gumbel distribution, see, e.g.,~\cite{embrechtsetal}.
\end{remark*}

\section*{Acknowledgments}
It is a pleasure for us to thank M.~Skriganov and
F.~Goethe who recommended us the book by E.~Grosswald~\cite{Grosswald}.

Both authors were partially supported by grant NSh-2251.2003.1 of the
President of Russian Federation.  The first author was also partially
supported by grant INTAS-03-51-5018; the second author was also
supported by the NWO postdoctoral fellowship.

\end{document}